\documentclass[11pt,prd,nofootinbib]{revtex4} 
\usepackage{graphicx} 
\usepackage{amsmath}
\usepackage{amsfonts,amsbsy}
\usepackage{amssymb}

\newcommand{\rb}{\mbox{\boldmath $b$}}

\newcommand{\be}{\begin{equation}}
\newcommand{\ee}{\end{equation}}

\begin{document}

\title{Heavy quark production in $pA$ collisions: the double parton scattering contribution.}
\author{E.R.  Cazaroto$^1$ , V.P. Gon\c{c}alves$^{2}$ and F.S. Navarra$^1$}

\affiliation{$^1$ Instituto de F\'{\i}sica, Universidade de S\~{a}o Paulo, CEP 05315-970 S\~{a}o Paulo, SP, Brazil\\
$^2$ Instituto de F\'{\i}sica e Matem\'atica, Universidade Federal de Pelotas, CEP 96010-900, 
Pelotas, RS, Brazil.}

\begin{abstract}
In this paper we estimate the double parton scattering (DPS) contribution for the heavy quark production in   $pA$ collisions at the LHC. The  cross sections for the charm and bottom production are estimated using  the dipole approach and taking into account the saturation effects, which are important for high energies and for the scattering with a large nucleus.  We compare the DPS contribution with the single parton scattering one and demonstrate that in the case of charm production both are similar in the kinematical range probed by the LHC. Predictions for the rapidity range analysed by the LHCb Collaboration are also presented. Our results indicate that the study of the DPS contribution for the heavy quark production in $pPb$ collisions at the LHC is feasible and can be useful to probe the main assumptions of the approach.
\end{abstract}

\maketitle

\section{Introduction}

In hadronic collisions at high energies the occurrence of multi-parton interactions (MPI) is a consequence of the high density of partons in the hadron 
wave functions. In this kinematic regime the huge number of gluons increases the probability that two or more hard gluon-gluon fusion in a single hadron -- 
hadron collision take place. The single gluon-gluon fusion in this kind of process is usually called Single Parton Scattering (SPS) and its contribution  
is in general the  dominant process in  perturbative QCD (pQCD) calculations. Recently, several theoretical and experimental studies have shown that 
Double Parton Scattering (DPS) processes cannot be neglected at LHC energies  (For  recent reviews see, e.g. Ref. \cite{review}).
In particular, the experimental results from the LHCb Collaboration on  four $D$ meson production in $pp$ collisions \cite{lhcb_jhep} indicate that the DPS contribution is non - negligible in the kinematical range considered. Besides accounting for a significant part of the cross section, the study of  DPS processes is 
also important for other reasons.  It can, for example, help us to understand the spatial structure of  hadrons \cite{diehl_jhep}, the multi-parton 
correlations in the hadronic wave function \cite{diehl_jhep,part_correl,salvini,blokcor,ww,osta} and is expected to help in the search for new physics 
(See, e.g., Ref. \cite{new_physics}).

One of the promising processes to probe the DPS mechanism is  heavy quark production. At high energies, this process probes the hadron wave function
at very small values of the Bjorken - $x$ and its cross section can be calculated perturbatively. This process is dominated by gluon - gluon scatterings 
and a large cross section is predicted at the LHC by the    single scattering mechanism. As a consequence of the large luminosity of small - $x$ gluons 
in the initial state,  we expect a significant contribution of the DPS mechanism to heavy quark production. This expectation has been confirmed by 
the analysis performed in Refs. \cite{antoni,russo1,dps_us} (See also Refs. \cite{outrosantoni,ww}). In particular, in Ref. \cite{dps_us} we have investigated  
the impact of saturation effects in  DPS production of heavy quarks. The results from Refs. \cite{antoni,dps_us} demonstrated that for   charm 
production in $pp$ collisions at LHC energies the double parton scattering contribution  becomes comparable with the single parton scattering one. Moreover, 
in Ref. \cite{dps_us} we also  demonstrated that the production of $c\bar{c}b\bar{b}$ contributes significantly to  bottom production.

Another possibility to probe the DPS mechanism is the analysis of different final states in nuclear collisions. The  studies performed in Ref. 
\cite{treleanistrik,dps_pa,dps_jpsi,blok,salvini} have shown that the DPS mechanism is strongly enhanced in $pA$ and $AA$ collisions. These studies 
encourage us  to extend our previous analysis to $pA$ collisions and investigate the DPS contribution to heavy quark production. In particular, we will 
estimate the magnitude of the DPS cross section for $pPb$ collisions    at $\sqrt{s_{NN}} = 5.02$ TeV, which can be measured at the LHC. As at 
small-$x$ and a large nucleus we expect a large contribution of  saturation effects to heavy quark production \cite{nos_npa}, we also include these 
effects in our calculations.

This paper is organized as follows. In the next Section  we present the basic assumptions and formulas derived in Refs. \cite{treleanistrik,dps_pa}, which 
we use to calculate the DPS cross sections for the heavy quark production in $pA$ collisions. In Section \ref{sec_results} we estimate the total cross 
section for the $c\bar{c} c\bar{c}$, $b\bar{b} b\bar{b}$ and $c\bar{c} b\bar{b}$ production for different nuclei and analyse its energy dependence. The 
DPS and SPS contributions are compared and the magnitude of the DPS contribution for $pPb$ collisions    at $\sqrt{s} = 5.02$ TeV is presented. Predictions 
for the kinematical range probed by the LHCb experiment also are show. Finally,  in Section \ref{sec_conclusions} we summarize our main conclusions.

\section{The formalism}
\label{sec_formalism}

Initially let us present a brief review of the formalism used to treat  single and double parton scattering in a generic hadron - hadron collision. In 
the case of a SPS process, we  assume that only one hard interaction occurs per collision. The basic idea, which justifies this approach, is that the 
probability of a hard  interaction in a collision is very small, which makes the 
probability of having two or more hard interactions in a collision highly suppressed with respect to the single interaction probability. As discussed in 
Refs. \cite{antoni,russo1,dps_us} such assumption is reasonable in the kinematical regime in which the flux of incoming partons is not very high.  However, 
at LHC energies there is a high probability of scattering of more than one pair of partons in the same hadron - hadron collision. Consequently, it is 
important to take into account the contribution of the DPS processes. Following the same factorization approximation assumed for processes with a single 
hard scattering, it is possible to derive the DPS contribution for the heavy 
quark cross section considering two independent hard parton sub-processes. It is given  by  (See, e.g. Ref. \cite{diehl_jhep})
\begin{eqnarray}
\sigma_{h_1 h_2 \rightarrow Q_1\bar{Q}_1Q_2\bar{Q}_2}^{DPS} = \left( \frac{m}{2} \right)  \int \Gamma_{h_1}^{gg} (x_1,x_2; \rb_1, \rb_2; \mu_1^2, \mu_2^2) 
\hat{\sigma}_{Q_1\bar{Q}_1}^{gg} (x_1,x_1^{\prime},\mu_1^2) \hat{\sigma}_{Q_2\bar{Q}_2}^{gg} (x_2,x_2^{\prime},\mu_2^2) \nonumber \\
\times  \Gamma_{h_2}^{gg} (x_1^{\prime},x_2^{\prime};\rb_1-\rb,\rb_2-\rb;\mu_1^2,\mu_2^2) dx_1 dx_2 dx_1^{\prime} dx_2^{\prime} d^2b_1 d^2b_ 2 d^2b \,\,,
\label{sigdps_geral}
\end{eqnarray}
where we assume that the quark-induced sub-processes can be disregarded at high energies,   $\Gamma_{h_i}^{gg} (x_1,x_2; \rb_1, \rb_2; \mu_1^2, \mu_2^2)$ are 
the two-gluon parton distribution functions which depend on the longitudinal momentum fractions $x_1$ and $x_2$, and on the transverse positions $\rb_1$ and 
$\rb_2$ of the two gluons undergoing  hard processes at the scales $\mu_1^2$ and $\mu_2^2$. The functions $\hat{\sigma}$ are the parton level sub-processes
 cross sections and $\rb$ is the impact parameter vector connecting the centres of the colliding hadrons in the transverse plane. Moreover, $m/2$ is a 
combinatorial factor which accounts  for  indistinguishable  and distinguishable  final states. For $Q_1 = Q_2$ one has $m=1$, while $m=2$ for $Q_1 \neq Q_2$.  
It is common in the literature to assume that the longitudinal and transverse components of the  double parton distributions can be decomposed   and that the 
longitudinal components can be expressed in terms of the product of two independent single parton distributions.   As in \cite{dps_us} we  will also assume 
the validity of these assumptions and consider that the DPS contribution to 
 the heavy quark cross section can be expressed in a simple generic form given by
\begin{eqnarray}
\sigma_{h_1 h_2 \rightarrow Q_1\bar{Q}_1Q_2\bar{Q}_2}^{DPS} = \left( \frac{m}{2} \right) \frac{\sigma^{SPS}_{h_1 h_2 \rightarrow Q_1\bar{Q}_1} 
\sigma^{SPS}_{h_1 h_2 \rightarrow Q_2\bar{Q}_2}}{\sigma_{eff}} \,\,,
\label{dps_fac}
\end{eqnarray}
where $\sigma_{eff}$ is a normalization cross section representing the effective transverse overlap of partonic interactions that produce the DPS process. Disregarding possible correlations between the properties of two partons inside a hadron , e.g. spins, colors, flavors, transverse and longitudinal momenta, it is possible to relate 
$\sigma_{eff}$ with the impact parameter integral of the overlap function $t({\bf b})$: $\sigma_{eff}=[\int d^2b\,t^2({\bf b})]^{-1}$, where 
$t({\bf b})=\int f({\bf b_1})f({\bf b_1-b})d^2b_1$ and  $f({\bf b})$ describes the transverse parton density in a given hadron. In general, it has been 
considered as a free parameter to be determined through fits to experimental $pp/p\bar{p}$ data. Recent results for multiple DPS measurements at Tevatron and LHC indicate that in order to reproduce the data we should have  $\sigma_{eff,pp}\approx 15 \pm 5$ mb \cite{review}. 
Eq. (\ref{dps_fac}), usually called ``pocket formula'', expresses  the DPS cross section as the product of two individual SPS cross sections assuming that 
the two SPS sub-processes are uncorrelated and do not interfere.  The validity of these strong assumptions at LHC and higher energies  is still  an open 
question, which has motivated several theoretical studies (See, e.g. Refs. \cite{diehl_jhep,ww}). However,   the  phenomenological analysis of different 
processes indicates that Eq. (\ref{dps_fac}) can be considered a reasonable first approximation for the treatment of  DPS processes.

\begin{figure}[t]
\begin{tabular}{cc}
\includegraphics[scale=0.5]{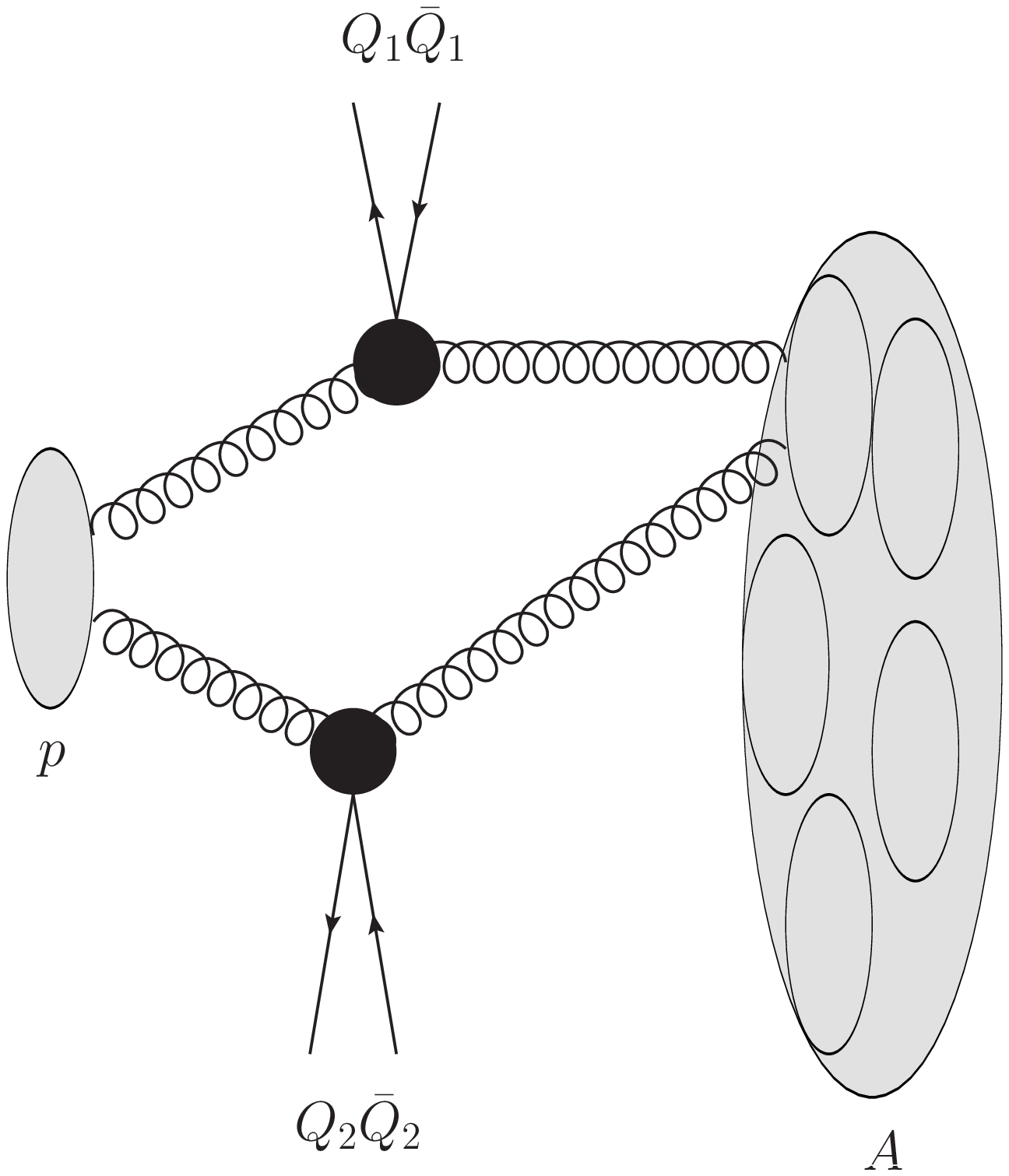}        
\hspace{1.cm}
\includegraphics[scale=0.5]{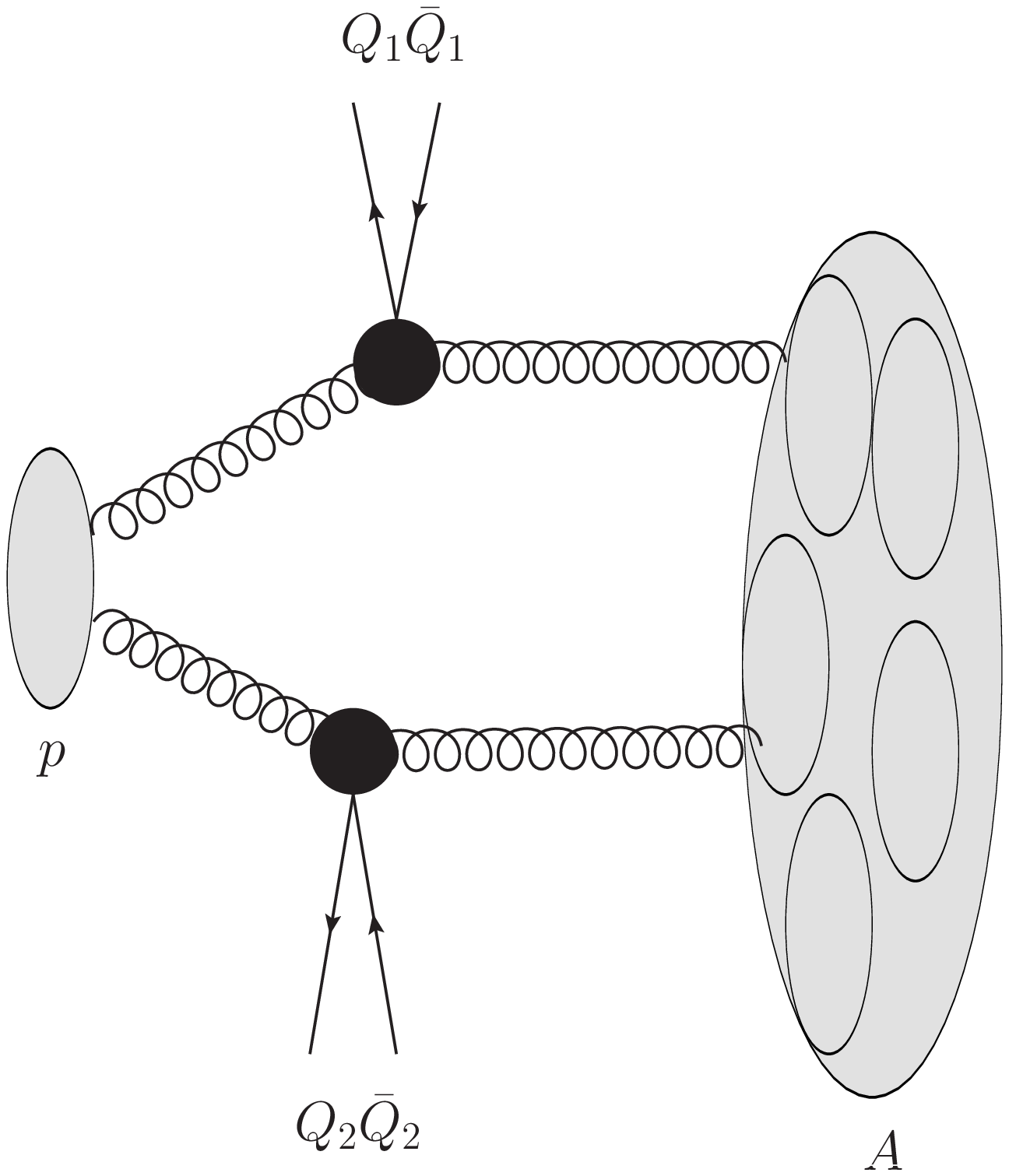}        
\end{tabular}
\caption{Heavy quark production through DPS in $pA$ collisions.  Left: Two gluons coming from the proton projectile scatter with two gluons 
coming from the same nucleon in the target nucleus; Right: Two gluons coming from the proton projectile scatter with two gluons coming from different 
nucleons in the  target  nucleus.}
\label{fig_DPS_pA}
\end{figure}

In order to extend  the treatment of DPS processes to proton - nucleus collisions we need to take into account that the parton flux associated to the 
nucleus is enhanced by a factor $\propto A$ and that in the interaction the two gluons associated to the proton can interact  
with two gluons coming from the same nucleon from the nucleus  or with two gluons coming from different nucleons from the nucleus. Both possibilities are 
represented in the left and right panels of the Fig  \ref{fig_DPS_pA}. Hereafter, we will denote the cross sections associated to these two contributions 
by   $\sigma_{pA}^{DPS,1}$ and  $\sigma_{pA}^{DPS,2}$, respectively.
A way to treat these contributions was proposed in  Ref. \cite{treleanistrik} and applied in Ref. \cite{dps_pa}  to  the production of same - sign $WW$  
in $pA$ collisions, which was suggested to be a signal for DPS. In what follows we extend the framework presented in Refs. 
\cite{treleanistrik,dps_pa} to the calculation of  heavy quark production. Following Refs. \cite{treleanistrik,dps_pa} we will assume that
$\sigma_{pA}^{DPS,1}$ can be estimated  scaling the proton - nucleon $pN$ cross section by the   number  $A$ of nucleons inside the nucleus, i.e. 
$\sigma_{pA}^{DPS,1}=A\cdot \sigma_{pN}^{DPS}$. Moreover, we will consider that 
$\sigma_{pA}^{DPS,2}$ can be estimated in terms of the DPS proton - nucleon cross section as follows:  
$\sigma_{pA}^{DPS,2}=\sigma_{pN}^{DPS}\cdot \sigma_{eff,pp}\cdot F_{pA}$. The quantity $F_{pA}$ can be expressed in terms of the  nuclear 
thickness function $T_{pA}$ as follows: $F_{pA}=[(A-1)/A]\,\int T^2_{pA}({\bf r})d^2r$, where  ${\bf r}$ is the 
impact parameter between the colliding proton and nucleus. As discussed in Ref. \cite{dps_pa}, the factor $(A-1)/A$ was introduced to take into account the difference between the number of nucleon pairs and the number of different nucleon pairs. Consequently, the final formula for the  DPS $pA$ cross section  is given by \cite{dps_pa}:
\begin{eqnarray}
\sigma^{DPS}_{pA \to ab}  =  \sigma_{pA}^{DPS,1} + \sigma_{pA}^{DPS,2} = A \sigma^{DPS}_{pN\to ab} \left[1 + \frac{1}{A} \sigma_{eff,\,pp}F_{pA}\right]  
\label{sig_dps_pa1}
\end{eqnarray}
which implies
\begin{eqnarray}
\sigma^{DPS}_{pA \to ab} & = &  \left(\frac{m}{2}\right) 
\frac{\sigma^{SPS}_{pN\to a} \cdot \sigma^{SPS}_{pN\to b}}{\sigma_{eff,\,pA}} \,\,,
\label{sig_dps_pa}
\end{eqnarray}
with the normalization effective cross section given by:
\begin{eqnarray}
\sigma_{eff,\,pA} = \frac{\sigma_{eff,\,pp}}{A+\sigma_{eff,\,pp}\,F_{pA}} \,\,.
\label{sig_eff_pa}
\end{eqnarray}
In the simplest approximation that the nucleus has a spherical form (with uniform nucleon density) of radius $R_A = r_0A^{1/3}$, and $r_0=1.25$ fm, the 
integral of the nuclear thickness factor becomes:
\begin{eqnarray}
F_{pA} = \frac{9A(A-1)}{8\pi R_A^2} \,.
\label{fpa}
\end{eqnarray}
The above equations one finds that $\sigma_{eff,\,pp}/\sigma_{eff,\,pA} \approx 3 \, A$ instead of the simple scale factor $A$ that one would naively expect. Moreover, this also implies that the $pPb$ DPS cross section is enhanced by a factor $3 A$ $( \approx 600)$ in comparison to the DPS contribution in $pp$ processes. 

The main input in the calculation of the DPS $pA$ cross section, Eq. (\ref{sig_dps_pa}), is the $pN$ cross section associated to the SPS process. As in 
our previous study \cite{dps_us}, we will estimate this quantity using the dipole approach, which  allows to easily include  saturation effects, which are expected to contribute significantly at the small values of $x$ 
probed in  heavy quark production at the LHC. This approach is expected to take into account of leading $\alpha_s \ln (1/x)$ corrections as well of QCD factorization breaking effects predicted to be present at large partonic densities \cite{kope_tarasov,raju,kovner}. Moreover,  the results presented in Ref. \cite{rauf} demonstrate the equivalence between the color dipole approach and the collinear one at low partonic densities, with the dipole predictions being similar to those obtained at next - to - leading order in the collinear formalism. Finally, as demonstrated in \cite{dps_us} (See also Ref. \cite{anna} for a recent analysis), this approach is able to describe the RHIC and LHC data. In the dipole 
approach the  
 total 
cross section for the process $pN \rightarrow Q\bar{Q} X$ is  given by \cite{Nikzak1,Nikzak2}:
\begin{equation}
\sigma (pN \rightarrow \{ Q\bar{Q} \}X) = 
2 \int _0 ^{-ln(2m_Q/ \sqrt{s})} dy  \, x_1 \, g_{p}(x_1,\mu _F) \,
\sigma (g N \rightarrow \{Q\bar{Q}\} X)
\label{sigtot}
\end{equation}
where $x_1g_{p}(x_1,\mu _F)$ is the projectile gluon distribution, the cross section  $\sigma (gN \rightarrow \{Q\bar{Q}\} X)$ describes  
heavy quark production in a gluon - nucleon interaction, $y$  is the rapidity of the pair and $\mu_F$ is the 
factorization  scale. The basic idea  of  this approach is that before interacting with the nucleon target $N$ a 
gluon is emitted by the projectile $p$, which fluctuates into  a color octet pair $Q\bar{Q}$. As in the low-$x$ regime the time of fluctuation is much larger
 than the time of interaction, and color dipoles with a defined transverse separation $\vec{\rho}$ are eigenstates of the interaction. 
The cross section for the process $g + N \rightarrow Q \bar{Q} X$ is given by:
\begin{equation}
\sigma(g N \rightarrow\{Q\bar{Q}\}X) = \int _0^1 d \alpha \int d^2\rho \,\, 
\vert \Psi _{g\rightarrow Q\bar{Q}} (\alpha,\rho)\vert ^2 
\,\, \sigma^{N} _{Q\bar{Q}g}(\alpha , \rho)
\label{sec1}
\end{equation}
where $\Psi _{g\rightarrow Q\bar{Q}}$ is the 
light-cone (LC) wave-function of the 
transition $g \rightarrow  Q \bar{Q} $ and $  \sigma^{N}_{Q\bar{Q}g}$  is the scattering cross section of a color neutral quark-antiquark-gluon system on the 
hadron target $N$ \cite{Nikzak1,Nikzak2,kope_tarasov,rauf}. As discussed in Ref. \cite{nos_npa,dps_us}, this cross section can be expressed in terms  of the 
dipole - proton cross section which is determined by the QCD dynamics at high energies and is probed in the deep inelastic scattering $ep$ processes studied at 
HERA. Eq. (\ref{sigtot}) can be directly generalized to describe the total cross section of heavy quark production in $pA$ collisions \cite{nos_npa} 
considering the fact that  color dipoles are eigenstates
of the interaction. Therefore the  ${Q\bar{Q}g}$-nucleus interaction can be expressed in terms of the
cross section on a nucleon target using the Glauber-Gribov formalism:
\begin{eqnarray}
\sigma^A_{Q \bar Q g}(x,\rho) = 2\, \int d^2 \rb \, \left\{ 1 - \exp \left[-\frac{1}{2}  \, 
\sigma^N_{Q \bar Q g}(x,\rho^2)  
\,T_A(\rb)\right] \right\} \,\,,
\label{enenuc}
\end{eqnarray}
where $T_A(\rb)$ is the 
nuclear profile 
function, which is obtained from a 3-parameter Fermi distribution for the nuclear
density normalized to $A$. As in our previous studies \cite{nos_npa,dps_us}, we will assume that the 
dipole - nucleon cross section can be described by  the phenomenological saturation model proposed by Golec-Biernat and Wusthoff  (GBW) in Ref.  \cite{gbw}. 
As demonstrated in Refs. \cite{nos_npa,dps_us}, the predictions for  heavy quark production using this simplified model are very similar to those obtained 
using as input the solution of the running coupling Balitsky - Kovchegov equation \cite{bkrunning}, which is the current state of the art of the  treatment of  
the non-linear and quantum effects in the hadron wave function. Moreover, following Ref. \cite{dps_us} we will  assume that $m_c = 1.5$ GeV, $m_b = 4.5$ GeV,  $\mu_F=2 m_Q$ and that $xg$ is 
given in terms of the leading - order  CTEQ10 parametrization \cite{cteq}, which allows to describe the RHIC and LHC data for the total cross sections. 
As verified in Refs. \cite{nos_npa,anna}, the predictions for the heavy quark production at LHC energies are strongly sensitive to these choices. In particular, for the charm production at $\sqrt{s_{NN}} = 13$ TeV, the upper and lower bound predictions for the total cross section can be different by $\approx 50 \%$ of the central one. 
Such large uncertainty is similar to that  present in the NLO collinear predictions (See Table 2 in Ref. \cite{anna}). In the next Section, we will discuss the implications of this uncertainty on our predictions for the DPS cross sections.

\begin{figure}[t]
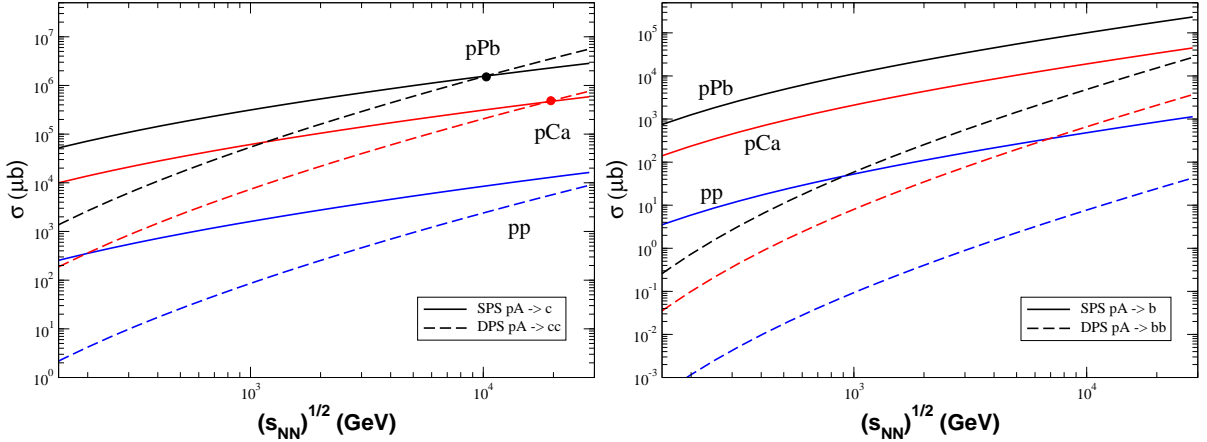

\begin{tabular}{cc}
\includegraphics[scale=0.33]{Fig_2a.eps}
\includegraphics[scale=0.33]{Fig_2b.eps}        
\end{tabular}
\caption{Central predictions for the energy dependence of the SPS and DPS cross sections for  charm (left panel)  and bottom (right panel) production in $pp$, $pCa$ and $pPb$ collisions. 
The SPS (DPS) predictions are represented by solid (dashed) lines. The current uncertainty in the predictions at high energies, associated to changes in the factorization and renormalization scales and heavy quark mass, is $\approx 40 \,(90) \%$ in the SPS (DPS) case. 
}
\label{f1_sps_dps_lhc}
\end{figure}

\section{Results and discussion}
\label{sec_results}
In what follows we will present our predictions for the integrated DPS $pA$ cross section of $c\bar{c} c\bar{c}$, $b\bar{b} b\bar{b}$ and $c\bar{c} b\bar{b}$ 
production. We will estimate $\sigma^{DPS}_{pA \to Q_1\bar{Q}_1Q_2\bar{Q}_2}$  considering the full rapidity range covered by the LHC as well as the rapidity 
range probed by the LHCb experiment ($2.0 < y < 4.5$). The single parton scattering cross section associated to the process $pN \rightarrow Q\bar{Q} X$ will be 
calculated using Eq. (\ref{sigtot}). For the case of a nuclear target, we will use Eq. (\ref{enenuc}) as input in our calculations. Moreover, 
we will assume that $\sigma_{eff,\,pp} = 15$ mb. Using Eq. (\ref{fpa}) we obtain that    $F_{pA} = 3.0 \, (28.1) $ mb$^{-1}$ for $A = 40 \, (208)$, which implies that  
$\sigma_{eff,\,pCa} = 170$ $\mu$b and $\sigma_{eff,\,pPb} = 23.8$ $\mu$b. Finally, in our analysis 
the contribution of the single parton scattering processes associated to the $gg \rightarrow Q_1\bar{Q}_1Q_2\bar{Q}_2$ diagram will  not be included, since the results presented in Ref. \cite{antoni_wolfgang} indicate that its magnitude is $\approx 2$ orders of magnitude smaller than the DPS contribution  in the kinematical range considered.

\begin{table}[t] 
\centering
\begin{tabular}{||c|c|c|c|c||} 
\hline 
\hline
Final state & Mechanism & $\sqrt{s_{NN}} = 2.76$ TeV & $\sqrt{s_{NN}} = 5.02$ TeV & $\sqrt{s_{NN}} = 8.8$ TeV\\
\hline 
\hline
$c\bar{c}$ & SPS  & 664 mb & 994 mb  &  1420 mb\\
$c\bar{c}c\bar{c}$ & DPS & 258 mb  & 602 mb  & 1280 mb \\
\hline
$b\bar{b}$ & SPS & 32 mb & 55 mb &  90 mb \\
$b\bar{b}b\bar{b}$ & DPS & 0.5 mb  & 1.5 mb  & 3.9 mb \\
\hline
\hline
\end{tabular}
\caption{Central predictions for the SPS and DPS contributions for  charm and bottom production in $pPb$ collisions at different  center - of - mass energies 
considering the full kinematical range covered by the LHC. The current uncertainty in these predictions, associated to changes in the factorization and renormalization scales and heavy quark mass, is $\approx 40 \,(90) \%$ in the SPS (DPS) case. } 
\label{tab1}
\end{table}

Initially let us analyse the nuclear dependence of SPS and DPS cross sections. As emphasized in the previous section, the DPS contribution in nuclear 
collisions is enhanced in comparison to $pp$ collisions. This can be observed in the results presented in Fig. \ref{f1_sps_dps_lhc}, where we show our 
predictions for  charm (left panel) and bottom (right panel) production. In the case of charm production we can see that the energy where the  SPS 
and DPS contributions becomes identical (indicated by a small  circle in the figure) decreases at larger values of $A$. We can see that for $A =$ 1 the equality takes place above the considered energy range, whereas 
$\sigma^{DPS}_{pA \to c\bar{c}c\bar{c}} = \sigma^{SPS}_{pA \to c\bar{c}}$  occurs at 
$\sqrt{s_{NN}}\, \approx \,$ 19.6 and 10.4 TeV for $A =$ 40 and 208, respectively. 
In the case of bottom production, the SPS and DPS contributions are identical only for energies beyond the range considered in the figure. 

In Table \ref{tab1} we present our predictions for the SPS and DPS cross sections for $pPb$ collisions at different center-of-mass energies. We  present only the results associated to the central predictions, obtained with the set of parameters, gluon PDF and dipole model, that allow us to describe the RHIC and LHC data for the SPS cross section in $pp$ collisions. Initially, let us discuss the SPS predictions for the charm and bottom production in $pPb$ collisions. We have verified that if the factorization and renormalization scales are modified within a factor of two and heavy quark mass are modified by $\pm 20\%$, the resulting predictions differ of the central values by a factor 1.4, which is similar to the uncertainty observed in the $pp$ case.
Our predictions can be compared with those presented in Ref. \cite{davidtriple}, which have estimated the cross sections using the collinear formalism at next - to - next leading order (NNLO) with  the nuclear modifications of the parton distributions being described by the EPS09 - NLO parametrization \cite{eps}. The results presented in Table \ref{tab1} are similar to  those presented in \cite{davidtriple}. In particular, our central predictions
are slightly larger than the central  results presented in \cite{davidtriple}. Such difference  is mainly associated to the fact that in our calculations we are assuming a smaller value for the heavy quark mass. As demonstrated in Ref. \cite{nos_npa}, our predictions are sensitive to the value of $m_Q$, with larger values reducing the magnitude of the cross section. It is important to emphasize  that saturation effects were not taking into account in Ref. \cite{davidtriple}. However, they consider the presence of shadowing effects at small - $x$, which also implies a reduction in the magnitude of the cross section. The origin of the shadowing effects included in EPS09 parametrization is still an open question, with the saturation physics being one of the possible alternatives (See e.g. Ref. \cite{review_armesto}).      

Let's now discuss the DPS predictions presented in Table \ref{tab1}. As in SPS case, we only present the central results. However, as these predictions 
were obtained using the pocket formula, Eq. (\ref{sig_dps_pa}), we can estimate the uncertainty present in the results. The results presented in Refs. \cite{nos_npa,anna} indicate that the current uncertainty in the dipole predictions for the charm and bottom production in $pp$ collisions is a factor $\approx 1.5$. As a consequence, 
we can estimate that the uncertainty present in the results shown in Table \ref{tab1} is of a factor $\approx 2$. This factor can be larger by approximately 20 \% due to current uncertainty in the value of $\sigma_{eff, pp}$. Although the normalization of the SPS and DPS cross sections can be modified by the current theoretical uncertainty, our results 
indicate that the DPS contribution for  charm production is non - negligible in the range of energies probed by the LHC in $pPb$ collisions, as it already was  
in $pp$ collisions \cite{dps_us}. In the case of the bottom production, our results indicate that $\sigma^{DPS}_{pA \to b\bar{b}b\bar{b}} \lesssim 0.05 \times \sigma^{SPS}_{pA \to b\bar{b}}$ at LHC energies.

\begin{figure}[t]
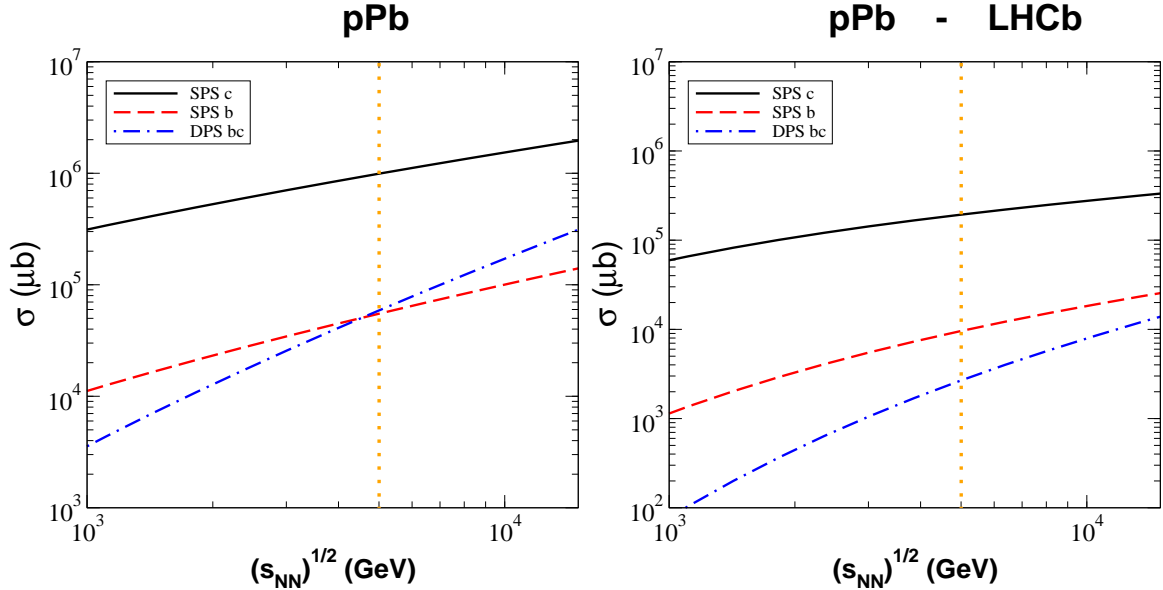

\begin{tabular}{cc}
\includegraphics[scale=0.45]{Fig_3a.eps}
\includegraphics[scale=0.45]{Fig_3b.eps}        
\end{tabular}
\caption{Comparison between the SPS predictions for charm (solid line) and bottom (dashed line) production and the DPS one for the production of the  
$b\bar{b}c\bar{c}$ final state (dot - dashed line) in $pPb$ collisions. In the left panel we present our predictions obtained considering the full 
rapidity range covered by the LHC, while in the right panel the cross sections were integrated over the rapidity range covered by the LHCb experiment 
($2<y<4.5$). Only the central predictions are presented, with the uncertainty in the results being similar to that indicated in the previous figures.}
\label{f3_dps_bc_lhc_lhcb}
\end{figure}

Another possible final state that can be produced considering the DPS mechanism is the $b\bar{b}c\bar{c}$ system, which can be generated when one gluon - gluon 
interaction creates a $b \bar{b}$ and the other a  $c \bar{c}$ pair. As demonstrated in Ref. \cite{dps_us}, the  DPS production of $b\bar{b}c\bar{c}$ can be 
responsible for approximately half of the total amount of bottom quarks produced in $pp$ collisions at the LHC. In what follows we will analyse how this conclusion 
is modified in $pPb$ collisions. In Fig. \ref{f3_dps_bc_lhc_lhcb} we compare the SPS production cross sections of $c\bar{c}$ and of $b\bar{b}$ pairs, denoted 
respectively by ``SPS c'' and ``SPS b'', with the DPS production cross section for the  $b\bar{b}c\bar{c}$ final state (denoted ``DPS bc'' in the figure). 
In the left panel we present our predictions obtained considering the full rapidity range covered by the LHC, while in the right panel the cross sections were 
integrated over the rapidity range covered by the LHCb experiment ($2<y<4.5$). The vertical dotted - lines indicates the center - of - mass energy of 5.02 TeV.
In the case that the cross sections are integrated over the full rapidity range, one has that the associated production of a $b\bar{b}$ with a $c\bar{c}$ becomes 
of the same order of the SPS production of a $b \bar{b}$ in $pPb$ collisions for energies of the order of 4 TeV, being dominant at larger energies. As expected, 
it occurs at smaller energies than in $pp$ collisions, where we have estimated that $b\bar{b}c\bar{c}$ and $b\bar{b}$ cross sections are similar only at  
$\sqrt{s_{NN}} \approx 10$ TeV. On the other hand, if the LHCb rapidity range is considered,  the  $b\bar{b}c\bar{c}$ cross section is a factor four  smaller than 
the $b\bar{b}$ one.

\begin{figure}[t]
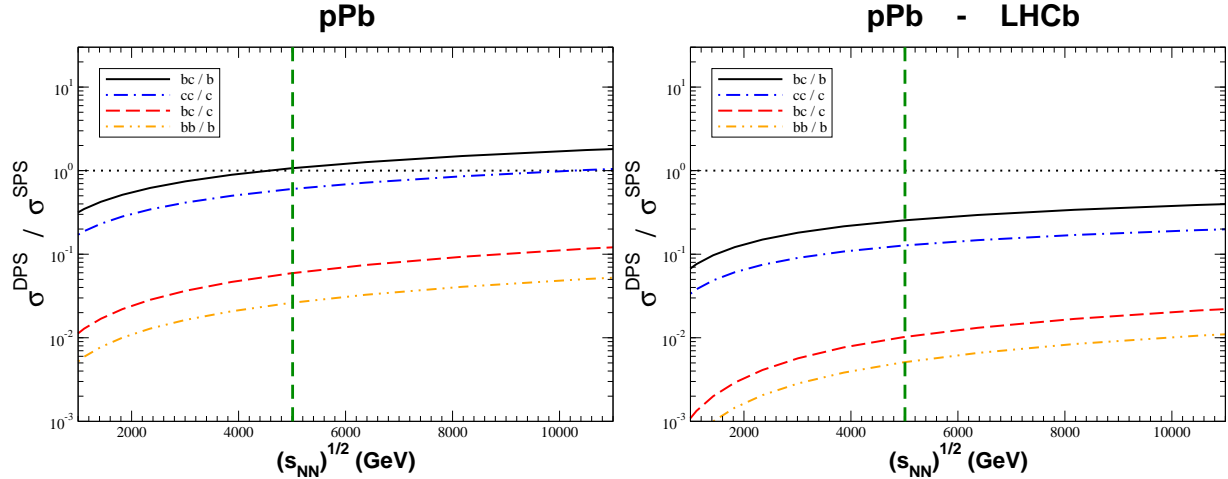

\begin{tabular}{cc}
\includegraphics[scale=0.33]{Fig_4a.eps}        
\includegraphics[scale=0.33]{Fig_4b.eps}        
\end{tabular}
\caption{Energy dependence of the ratio between the  DPS and  SPS cross sections for different combinations of final states. Left panel: Cross sections 
integrated over the full LHC rapidity range. Right panel: The cross sections are integrated over the rapidity range of the LHCb experiment ($2<y<4.5$).}
\label{f2_ratios_lhc_lhcb}
\end{figure}

In order to obtain a more precise estimate of the DPS contributions relative to the SPS ones, in Fig. \ref{f2_ratios_lhc_lhcb} we present the energy 
dependence of the ratio $\sigma^{DPS} / \sigma^{SPS}$ for different final states. We denote by  ``$bc/b$''  the ratio between the DPS production of 
$b\bar{b}c\bar{c}$ final state and the SPS production of $b\bar{b}$ pair, with analogous notation for the other combinations. In the left panel we present 
the predictions for the full LHC rapidity range, while in the right panel we integrated over the rapidity range covered by the LHCb experiment. The 
vertical dashed line indicates $\sqrt{s_{NN}} = 5.02$ TeV. 
Our results for the full rapidity range indicate that the ratios 
``$bc/b$'' and ``$cc/c$'' are of order of unity in $pPb$ collisions at $\sqrt{s_{NN}} = 5.02$ TeV, while the ratios ``$bc/c$'' and ``$bb/b$'' are smaller 
than 0.05. In contrast, all ratios are smaller than 0.3 in the LHCb rapidity range. In Ref. \cite{dps_us} we estimated these same ratios for $pp$ 
collisions. Comparing the above results obtained for $pPb$ collisions with those presented in Fig. 4 of Ref. \cite{dps_us}, we have that  these are 
considerably greater. Therefore, even at the rapidity range of the LHCb,  heavy quark production in DPS processes is more likely to be experimentally 
detected in $pPb$ collisions than  in $pp$ collisions. As pointed in Ref. 
\cite{dps_pa}, this can be useful to constrain the value of $\sigma_{eff,\,pp}$, since $F_{pA}$ is reasonably well determined from the nuclear geometry 
[See Eqs. (\ref{sig_dps_pa}) and (\ref{sig_eff_pa})].

\section{Conclusion}
\label{sec_conclusions}
Recent experimental and theoretical studies of different final states that can be produced in $pp$ collisions at the LHC have demonstrated that the 
contribution of  double parton scattering processes can be non - negligible and should be taken into account. Such contribution becomes large at high 
energies due to the large parton luminosity  in the initial state and is enhanced in  nuclear collisions. In this paper we have extended 
our previous study of DPS production of heavy quarks in $pp$ collisions to  $pA$ collisions. We have used the dipole approach and we have taken into account 
the saturation effects which are expected to be important for small $x$ and large nuclei. We estimated the $A$ dependence of the SPS and DPS cross sections 
and demonstrated that the DPS contribution for  charm production is similar to the SPS one for $pPb$ collisions at $\sqrt{s_{NN}} = 5.02$ TeV and dominates 
at larger energies. Additionally, we have shown that the associated production of a $b \bar{b}$ with a $c \bar{c}$ has a cross section similar to the SPS cross 
section for the production of a $b \bar{b}$. Our results indicate that the analysis of the $c \bar{c}c \bar{c}$
and $b \bar{b}c \bar{c}$ final states in $pPb$ collisions at the LHC can be useful to constrain the double parton scattering mechanism.

\section*{ACKNOWLEDGEMENTS}
This work was  partially financed by the Brazilian funding
agencies FAPESP, CNPq, CAPES, FAPERGS and  INCT-FNA (process number 
464898/2014-5).

\end{document}